\documentclass[12pt]{article}
\begin{document}
\sf
\begin{center}
   \vskip 2em
{\LARGE \sf  Laplace pressure as a surface stress in fluid vesicles}
\vskip 3em
 {\large \sf  Jemal Guven \\[2em]}
\em{
Instituto de Ciencias Nucleares,
 Universidad Nacional Aut\'onoma de M\'exico\\
 Apdo. Postal 70-543, 04510 M\'exico, DF, MEXICO\\[1em]
}
\end{center}
 \vskip 1em

\begin{abstract}
Consider a surface, enclosing a fixed volume, 
described by a free-energy depending only on the local geometry; for example, 
the Canham-Helfrich energy quadratic in the mean curvature describes a fluid membrane. 
The stress at any point on the surface is determined completely by  geometry. 
In equilibrium, its divergence is proportional to the Laplace pressure, 
normal to the surface, maintaining the constraint on the volume.
It is shown that this source itself can be expressed as the divergence  of
a position-dependent surface stress. As a consequence,  the  equilibrium can be 
described in terms of a conserved  {\it effective} surface stress. 
Various non-trivial geometrical consequences of this identification are explored.
In a cylindrical geometry, the cross-section can be viewed as a closed planar Euler 
elastic curve. With respect to an appropriate centre 
the effective stress itself vanishes; this provides    
a remarkably simple relationship between the curvature and the position
along the loop. In two or higher dimensions, it is shown that  
the only geometry consistent with the vanishing of the effective stress 
is spherical. It is argued that the appropriate generalization of the loop result will 
involve {\it null} stresses. 
\end{abstract}
\today \vskip 1em PACS: 04.60.Ds \vskip 3em

\section{\sf Introduction}

The energy of a soap bubble is proportional to its surface area.
The bubble is round because this is the shape of the surface, enclosing a 
fixed volume, which minimizes area \cite{Kenmotsu}. 
The shape of a closed surface described by
an energy which depends in some more general way on the geometry
need not be spherical. A 
striking example is provided by biology; 
the fluid membranes in our cells are described with uncanny accuracy,
on optical scales,  by the Canham-Helfrich (or Willmore) energy quadratic 
in the mean curvature of the membrane surface \cite{CanHel,Willmore}. 
In addition to the volume the area is also fixed. There may also be 
other constraints. Under these circumstances,
determining the equilibrium shape, never mind 
addressing questions of stability, 
becomes a sutble problem. We know, however, 
thanks to a series of landmark numerical studies 
in the nineties that very different equilibrium shapes may be 
consistent with given values of these constraints and that large 
changes in shape may be brought about by tweaking the constraints slightly.
This work is nicely summarized  
in references \cite{SackLip,Reviews} and \cite{Nonmed}. 
Representative examples of recent work are provided by \cite{Wor} and \cite{Zit}.
Our current theoretical framework, by constrast, remains 
woefully undeveloped. 
Even if we artificially limit ourselves  to axially symmetric configurations, 
where it is possible to boast a level of analytical control, 
there is a significant  discrepancy between the wealth of numerical data that 
has accumulated and our current ability to interpret the patterns within it  
 --- even qualitatively --- on the basis of the underlying model. 
Some new element needs to be introduced 
into our theoretical framework if this imbalance is to be set right.

It is well known that is it possible to describe the equilibrium shape of the surface
in terms of the conservation of a stress tensor. This problem was first approached,
a la Gibbs, 
thermodynamically \cite{Evans1,Kral};
such an approach, however, never really exploits the fact that 
the energy depends only on geometrical degrees of freedom; when the scaffolding is removed,
what is left is geometry. The appropriate setting has
been described in \cite{Stress,auxil}. Recently, Lomholt 
and Miao have shown how the two approaches can be reconciled \cite{LomMiao}. 
From the geometrical point of view, the existence of the conservation 
law is a consequence of the fact 
that the free energy is invariant under translations. 
Noether's theorem identifies the stress as the 
conserved current associated with this invariance. Furthermore,
the invariance of the geometrical free-energy   
under reparametrizations of the surface implies that 
the stress is determined completely by the surface
curvature and its derivatives. Knowledge of the local membrane 
geometry will always be sufficient to  reconstruct the stress underpinning it. 
This contrasts with the behavior in an elastic solid 
where the stress is proportional to 
first derivatives and the geometry itself is not always enough to 
determine the state of stress \cite{Ruso}. 

Internal constraints on the geometry will add stress to the membrane.
In the case of a fluid membrane, as we have mentioned,
the area is fixed; the two layers of a bilayer may also have different areas.
The latter asymmetry is captured by a constraint or a penalty on the
total mean extrinsic curvature of the surface \cite{Sve.Zek:89}. 
These constraints are accommodated using
Lagrange multipliers. 

Whereas surface constraints add to the stress due to bending energy, 
a constraint on the enclosed volume acts as an external force on the 
membrane. This is  the well known Laplace pressure counteracting the tension in a soap bubble 
which acts normally to the surface. 
In this note, it will be shown how to incorporate this external pressure 
into a divergence-free effective surface stress. This makes it possible to 
treat the volume constraint on an equal footing with its surface counterparts.
Geometrically, this is because the surface normal along which the 
Laplace force acts  can always be expressed  as a {\it surface} divergence: a fact
which is itself a direct consequence of the translation 
invariance of volumes. Curiously, the translational invariance of the stress gets broken 
in the construction; the effective stress depends explicitly on the 
position.  

The interesting point is that this unprepossessing mathematical identity 
permits us to look at the equilibrium in a new light.
We first examine a model describing 
cylindrical geometries in terms of this effective stress tensor. 
The cross-section of a cylinder perpendicular to its axis  
is a closed planar loop; the free-energy on this geometry 
describes one of Euler's elastica subject to 
a constraint on the enclosed area on the plane. This system was
studied in some detail in \cite{ACCG} where it was dubbed, with a
nod to Euler, {\it Elastica Hypoarealis}.
While it is well known that the Euler-Lagrange equation can be integrated 
to provide a quadrature for the curvature, this system 
exhibits an additional level of integrability
which is not evident in its Euler-Lagrange equation:
the curvature at any point detemines completely the geometry of the loop at that point. 
Such an identity is surprising because one would 
expect to require two integrations of the curvature to reconstruct the
corresponding loop whereas none is needed. In this paper, it  will be shown that this
is no accident. In fact, it is because the conservation law can be integrated:
modulo a translation centering the loop at the origin, the effective stress 
vanishes everywhere on the loop. 
The identity between curvature and position 
pointed out in \cite{ACCG} follows 
as an immediate consequence. Given the curvature, it is possible to trace 
the loop. One has not, of course, integrated the shape equation: one still needs 
to integrate the quadrature for the 
curvature as a function of arc-length in terms of elliptic integrals. But one does  
possess a relationship between curvature and the configuration which has 
sidestepped completely the need to evaluate these integrals.

The construction of the identity for a loop is loosely analogous to the 
Wulff construction of the equilibrium shape of a crystal from a 
knowledge of its anisotropic surface tension \cite{Wulff}. 
Is there an analogue for genuine two-dimensional membranes? 
Just as the equilibrium loop possesses a centre of `gravity',
there also will be one for a membrane. 
The relationship between curvature and the location with respect to 
this centre originates in the explicit spatial dependence of the effective stress.
In a spherical equilibrium, it is simple to check that
this stress vanishes when the origin is translated to the centre of the sphere. 
This is, in itself, a non-trivial identity 
concerning the balance of internal forces in the spherical equilibrium.
It is, however, also simple to prove in the case of a fluid membrane 
that spheres are the only compact  geometries possessing this property.\footnote 
{\sf In the case of the Clifford torus, one can check explicitly that 
the effective stress is non-vanishing.} One does not appear to have new to say about the
shape of membranes.

There is, of course, a good reason why a straightforward analogue of the 
loop identities does not exist in higher dimensions.
On the loop, a divergence-free (one-dimensional) effective stress 
is trivially constant, and this constant can be set to zero by 
centering the loop appropriately. 
In two-dimensions, however, there is enormously more freedom.
The analogue of a constant vector in the plane is now a 
{\it null} stress, divergence-free by construction. The null stresses on a surface
are in one-to-one correspondence with spatial vector fields. Furthermore, consistency with 
the symmetries of the fluid membrane stress tensor will place a constraint on this 
vector field; one finds that the admissible vector fields are 
the generators of area-preserving deformations of the surface.  Such 
stresses cannot generally be eliminated by a 
simple translation. As a consequence, the higher dimensional generalization of 
the result for loops will involve null stresses. It is too early to 
know if this going to be tractable. However, given the enormous stakes involved (and not
just for fluid membrane physics) this is certainly worth exploring. 

\section{\sf First order variations, Euler-Lagrange derivatives, translation
invariance and the stress}

The membrane will be described as a parametrized surface,
$\xi^a \to {\bf X} (\xi^a)$. The 
${\bf X}= (X^1,X^2,X^3)$ are three functions of two parameters $\xi^a$, $a=1,2$.
The metric tensor induced on this surface 
is given by $g_{ab}= {\bf e}_a\cdot {\bf e}_{b}$; 
the corresponding extrinsic curvature is
$K_{ab}= {\bf e}_a\cdot \partial_b\, {\bf n}$, 
where ${\bf e}_a= \partial_a{\bf
X}$, $a=1,2$ are the two tangent vectors to the surface 
and ${\bf n}$ is the unit normal.
$\nabla_a$ represents  the covariant derivative compatible with
$g_{ab}$. The Gauss-Weingarten equations, describing how these
vectors change as one moves across the surface, 
are $\nabla_a {\bf e}_b = - K_{ab}\, {\bf n}$ and
$\nabla_a {\bf n} = K_a{}^b\, {\bf e}_b$. Indices are raised and lowered with the metric.
When one looks at cylinders, the {\it surface} of interest 
will be a planar curve. Higher dimensional generalizations are 
also of potential interest; because it is straightforward 
to frame the discussion in terms of $D$-dimensional surfaces this is what will be done unless
explicitly indicated otherwise. Geometrical background material is provided in \cite{Carmo,Spivak}.
A nice tutotial tailored to membrane geometry is provided by reference \cite{Deserno}.

Consider any geometrical integral of the form
\begin{equation}
H [{\bf X}]= \int dA \, {\cal H}(g_{ab}, K_{ab})\,,
\label{eq:Hdef}
\end{equation}
defined on a  patch of surface $\Sigma$ with a boundary $\partial\Sigma$. 
$H$ will represent the free energy of the surface. ${\cal H}$ is a function of the
metric and the extrinsic curvature. Thus $H$ will be a functional of ${\bf X}$ and it 
will be invariant under reparametrizations and Euclidean motions of
the ambient space. The area element induced on the surface is $dA
=\sqrt{{\rm det}\,g_{ab}} d^2\xi$.
For simplicity, gradients of $K_{ab}$ are not considered;
they have, however, been considered in some detail elsewhere 
\cite{JG05}.

Recall how the stress tensor corresponding to $H$ gets constructed.
This involves examining the response of the free energy to a small deformation of the surface.
This deformation is described by the  infinitesimal change in the
embedding functions: 
${\bf X} \to {\bf X} + \delta {\bf X}$.
The response of $H$ decomposes naturally into two parts: 
a bulk response as well as  a boundary one; the latter is a surface divergence originating in
derivatives of the deformation in the bulk which have been consigned to the boundary.
One finds that
\begin{eqnarray}
\delta H &=& \int dA \, {\cal  E}\,{\bf n} \cdot\delta {\bf X} -
 \int dA \, \nabla_a [{\bf f}^a \cdot \delta {\bf X}]
\nonumber\\
&&   -\int dA\, \nabla_a \left[{\cal H}^{ab} \cdot
\nabla_b \,\delta {\bf X}\right] 
\,. \label{eq:delH0}
\end{eqnarray}
Here ${\cal E}{\bf n}$ is the Euler-Lagrange derivative
of $H$ with respect to ${\bf X}$; because of reparametrization invariance of
$H$, only the projection of $\delta {\bf X}$ onto the normal contributes to the bulk variation.
${\bf f}^a$ is the surface stress tensor associated with $H$ and 
${\cal H}^{ab}= \partial {\cal H}/\partial K_{ab}$. 
It has been shown elsewhere \cite{Stress,auxil} that
\begin{equation}
{\cal E}= -\nabla_a \nabla_b {\cal H}^{ab} + K_{ac} K^c{}_b {\cal
H}^{ab} + {\cal H}K\,;
\label{eq:ELdef}
\end{equation}
and 
\begin{equation}
{\bf f}^a = 
( {\cal H}^{ac}K_{c}{}^b -{\cal H}g^{ab}) {\bf
e}_b - \nabla_b {\cal H}^{ab} {\bf n} \,.
\label{eq:fdef}
\end{equation}
When $H$ is translationally invariant
it is possible to rewrite its  Euler-Lagrange derivative as  a surface
divergence. The argument of this divergence is the stress. This follows 
from Eq.(\ref{eq:delH0}):
consider the effect on the surface of a translation, $\delta {\bf
X}={\bf a}$, where ${\bf a}$ is a constant vector. The derivative
term appearing on the second line in Eq.(\ref{eq:delH0}) vanishes;
translational invariance,  $\delta H=0$, then implies that
\begin{equation}
{\bf a}\cdot \int dA \,[ {\cal  E}\,{\bf n}  -
 \nabla_a {\bf f}^a ]=0\,.
\label{delHt}
\end{equation}
Because the region of integration itself is arbitrary, the local statement 
\begin{equation}
{\cal  E} {\bf n} - \nabla_a {\bf f}^a =0 \label{eq:Ef}
\end{equation}
follows; ${\cal E}\,{\bf n}$ is a 
surface divergence. This is a useful identity to remember.

In particular, a fluid membrane is described by the
Canham-Helfrich energy
\begin{equation}
H_0 = {1\over2}\int dA\, K^2\,,
\end{equation}
where $K= g^{ab} K_{ab}$. It's mathematical properties, notably
its conformal invariance, were examined long beforehand \cite{Willmore}. 
Suppose that both the area $A$ and the integrated mean curvature 
\begin{equation}
M=\int dA\, K
\end{equation}
are fixed. 
The free-energy describing the surface is then given by
\begin{equation}
H= H_0 + \beta (M -M_0) + \sigma (A- A_0) \,,
\label{eq:Hfluid}
\end{equation}
where $\beta$ and $\sigma$ are two Lagrange multipliers enforcing the constraints. The 
constants  $M_0,A_0$ are the (for our purposes) irrelevant 
fixed valued of $M$ and $A$. 

\section{\sf Equilibrium with constrained volume}

Consider now the problem of determining the equilibrium of a
closed membrane subject to a constraint on the
enclosed volume. Introduce a new energy functional $\bar H$ related to the 
surface free energy $H$ by 
\begin{equation}
\bar H= H - P (V - V_0)\,.
\label{eq:HV}
\end{equation}
The constant Laplace pressure $P$ is a Lagrange multiplier 
forcing the volume to assume some fixed value, 
$V_0$. In equilibrium,  the Euler-Lagrange equation ${\cal E}=P$ is satisfied
(${\cal E}$ is the Euler-Lagrange derivative of $H$).
Using Eq.(\ref{eq:Ef}),
this equation can be cast as a conservation law:
\begin{equation}
\nabla_a  {\bf f}^a =P\, {\bf n}   \,.\label{eq:conserve}
\end{equation}
The external pressure provides a source for the stress in the membrane.
It is always possible, however,  as we will now 
demonstrate to incorporate this particular 
source into an effective conserved surface stress. 

\section{\sf Cone volume and the effective stress}

The stress was identified in section 2 by 
examining the boundary behaviour of the energy under deformation.
One way to place the volume $V$ appearing in the constrained energy given by 
Eq.(\ref{eq:HV}) on an equal footing with the surface invariants appearing in $H$ 
is to associate a  notion of volume with patches  of the surface.
But this is easy to do: fix an origin, and construct the cone ${\cal C}$ with 
a base which consists of the patch and its apex positioned at the origin. 
By examining how the volume of this cone behaves under deformations of the surface
patch, we will see that it is possible to associate an effective surface stress tensor with the volume
term appearing in Eq.(\ref{eq:HV}). It it then straightforward  to recast the 
equilibrium condition under the external Laplace pressure in terms of 
a divergence-free effective surface stress. 

First consider a closed surface. The volume of space it encloses is given by
\begin{equation}
V ={1\over D+1}\int dA \, {\bf n}\cdot {\bf X}\,.
\label{eq:defV}
\end{equation}
This is a direct consequence of the elementary Euclidean space identity for the position vector
${\bf x}$, ${\bf div}\cdot {\bf x}=D+1$ and the use of the divergence theorem. 
While it is possible to express $V$ as a surface
integral, it is {\it not} possible to express it 
in the form (\ref{eq:Hdef}). Despite this, as we will now show, an 
analogue of Eq.(\ref{eq:delH0}) does hold.

Suppose that the integral on the rhs of Eq.(\ref{eq:defV})
is restricted to some surface patch. It is
easy to see that the integral then represents the volume of the cone
${\cal C}$ that was described at the beginning 
of this section (see \cite{Spivak}).

Consider now the response of the volume of this cone to a
deformation of the surface ${\bf X}\to {\bf X}+\delta {\bf X}$.
This will point us towards an extremely useful identity for the normal vector.
One finds that
\begin{equation}
\delta V = \int dA \,{\bf n} \cdot\delta {\bf X} - {1\over D+1} \int dA \,
\nabla_a \left({\bf f}_0^a \cdot \delta {\bf X}\right)\,,
\label{eq:delV1}
\end{equation}
where
\begin{equation}
{\bf f}_0^a =
({\bf e}^a\cdot {\bf X}) \, {\bf n}
-({\bf n}\cdot {\bf X}) \, {\bf e}^a 
=
{\bf X}\times ({\bf n}\times {\bf e}^a)\,.
\label{eq:f0def}
\end{equation}
The second expression appearing on the rhs of Eq.(\ref{eq:f0def})
is valid for two-dimensional surfaces.
The derivation is straightforward; the details are 
provided for the interested reader in an appendix.
In particular, the well known result that
the Euler-Lagrange derivative of volume with respect to ${\bf X}$  is ${\bf n}$ 
is reproduced.
On a closed surface, the second term in
Eq.(\ref{eq:delV1}) vanishes so that the familiar area times normal 
displacement formula for the shell volume $\delta V$ follows. 

The stress-like object ${\bf f}_0^a$ we have introduced has a very nice property:
its surface divergence is proportional to ${\bf n}$:
\begin{equation}
{\bf n} = {1\over D} \nabla_a \,{\bf f}_0^a
\label{eq:ndiv}
\,.
\end{equation}
Curiously, the fact that ${\bf n}$ is a surface divergence appears to have gone unnoticed. 
The value of Eq.(\ref{eq:ndiv}) is that it provides  a relationship analogous 
to Eq.(\ref{eq:Ef}) for the volume with 
${\cal E}=1$:  ${\bf f}_0^a/D$ is identified as the 
effective surface stress associated with the volume term in the energy. 
The derivation of Eq.(\ref{eq:ndiv}) is elementary:
note that the curvature terms, which originate in
the Gauss-Weingarten equations for
the surface, cancel when the two terms are summed.
This is consistent with the fact that the normal depends only on  
first derivatives of ${\bf X}$. 

A consequence of this identification is that the Laplace force on a patch of 
surface can be cast as a closed line integral along its boundary:  
\begin{equation}
P \int dA \, {\bf n} = 
{P\over 2} \oint ds \,
({\bf X} \times {\bf t})
\,.
\label{eq:LapF}
\end{equation}
Here ${\bf t}$ is tangent to the
boundary curve and $s$ is arc-length. 
\footnote {\sf We have used the fact that
for right-handed triad $\{{\bf l},{\bf t},{\bf n}\}$, 
${\bf t}= {\bf n}\times {\bf l}$, where
${\bf l}= l^a {\bf e}_a$ is its outward pointing normal to the curve lying on the
surface. Any two orthonormal vectors specify a third; in particular,
the two normals ${\bf n}$ and ${\bf l}$ to the curve fix
its tangent.} 
In this form, it is clear that the Laplace force on the 
patch  depends only on the boundary geometry, a fact that is not 
at all obvious when cast as an integral over the patch.

Note that
the effective stress is {\it not} ${\bf f}_0^a/(D+1)$ as 
a casual comparison  of Eq.(\ref{eq:delV1}) with Eq.(\ref{eq:delH0}) might suggest;
this is because the cone volume, unlike the volume of a closed surface, is
not translationally invariant if the apex of the cone is fixed.
Indeed under a constant translation ${\bf a}$, 
${\bf X}\to
{\bf X}+ {\bf a}$, and it is clear from Eq.(\ref{eq:defV}) that 
\begin{equation}
\delta V={1\over D+1}\,{\bf a}\cdot  \int dA \, {\bf n}\,,
\end{equation}
a result consistent 
with Eq.(\ref{eq:delV1}) with the replacement of $\delta{\bf X}$ by ${\bf a}$.

The structure of the boundary term in 
Eq.(\ref{eq:delV1}) is of interest. 
On a two-dimensional surface,
it is possible to cast this equation  in the form
\begin{equation}
\delta V = \int dA \,{\bf n} \cdot\delta {\bf X} + {1\over 3} \oint ds \,
({\bf t} \times {\bf X})\cdot \delta {\bf X}
\,.
\label{eq:delV2}
\end{equation}
In this form, it is clear that the boundary term 
depends only on the boundary curve and its deformation;
in particular, it is independent of what the patch gets up to in its interior.

\section{\sf Laplace pressure as equivalent surface
stress}

The identification (\ref{eq:ndiv}) 
makes it possible to recast the 
equilibrium condition under an external Laplace pressure in terms of 
a divergence-free effective surface stress. Introduce the {\it effective} stress tensor, 
\begin{equation} \bar {\bf f}^a = {\bf f}^a -
{P\over D} {\bf f}_0^a\,.
\end{equation}
where ${\bf f}_0^a$ is given by 
Eq.(\ref{eq:f0def}).
Using Eq.(\ref{eq:ndiv}), it is possible to recast the equilibrium condition
Eq.(\ref{eq:conserve}) in the form
\begin{equation}
\nabla_a\, \bar {\bf f}^a =0\,.
\label{eq:nabbar}
\end{equation}
$\bar {\bf f}^a$ unlike ${\bf f}^a$ is divergence-free in equilibrium. 
In an isolated membrane, 
it follows from Eq.(\ref{eq:nabbar}) that
\begin{equation} 
\oint ds \,l_a \bar {\bf f}^a =0
\label{eq:gconserv}
\end{equation}
along any contractible curve which lies on the surface.
With the original stress tensor, 
the corresponding integral depends on the  
behaviour of the normal vector throughout the enclosed patch. 
Using the effective stress tensor, the volume constraint is treated like the 
other surface constraints.

If an external force acts on the patch of membrane enclosed by the curve,
the line integral appearing in Eq.(\ref{eq:gconserv}) will determine this force. 
In this context, it is often possible to deform the contour appropriately 
to take advantage of any symmetry the configuration might possess.
This approach was developed in \cite{MDG} to examine 
surface mediated interactions. Its generalization to incorporate 
a pressure difference across the membrane is also possible.

The effective stress $\bar {\bf f}^a$  has one very striking feature  
we have yet to comment upon.
Unlike ${\bf f}^a$, $\bar {\bf f}^a$ 
is {\it not} translationally invariant. Its 
origin is the explicit ${\bf X}$ dependence appearing 
in ${\bf f}_0^a$ given by Eq.(\ref{eq:f0def}).
Under a translation ${\bf X}\to {\bf X}+{\bf a}$, 
one finds that
${\bf f}_0^a \to {\bf f}_0^a + {\bf h}_0^a$,
where
\begin{equation}
{\bf h}_0^a=  
{\bf a}\times ({\bf n}\times {\bf e}^a)\,.
\label{eq:f1def}
\end{equation}
This shift should leave no physical trace. 
In the next section, this issue will be examined in the appropriate
setting.

\section{\sf Null stresses}
\label{sec:amb}

It is clear that there is an inherent ambiguity in the definition of any 
divergence-free effective stress
$\bar {\bf f}^a$; this is because one is always  free to add to $\bar {\bf f}^a$
any stress of the form
\begin{equation}
{\bf h}^a= \nabla_b {\bf A}^{ab}\,,
\label{eq:hdefg}
\end{equation}
involving  an antisymmetric 
potential ${\bf A}^{ab}= -{\bf A}^{ba}$. 
(In the case of a curve, this ambiguity degenerates into the freedom to add a constant vector.) 
It is straightforward to show that for ${\bf h}^a$ constructed according to 
Eq.(\ref{eq:hdefg}), one has $\nabla_a {\bf h}^a=0$ or
${\bf h}^a$ is conserved.
We will refer to such an 
${\bf h}^a$ as a {\it null} stress. In the language of differential forms
${\bf h}^a$ is exact \cite{Flanders}.

On a two-dimensional surface, the anti-symmetric potential ${\bf A}^{ab}$ 
factorizes: it must be a product of 
the two-dimensional Levi-Civita anti-symmetric tensor $\epsilon^{ab}$
and a space vector potential ${\bf A}$:
${\bf A}^{ab}= \epsilon^{ab} \,{\bf A}$. 
Because $\epsilon^{ab}$ is covariantly 
conserved ($\nabla_a \epsilon_{bc}=0$), it is possible to express
the corresponding null stress in the simple form \cite{Stress}
\begin{equation}
{\bf h}^a= \epsilon^{ab} \nabla_b {\bf A}
\,.
\label{eq:hdef}
\end{equation}
How various ambiguities in the definition of ${\bf f}^a$ get captured 
in ${\bf h}^a$ are discussed very nicely in \cite{LomMiao}. 

Suppose now that we add an ${\bf h}^a$ constructed this way to the stress;
its contribution to the force on any patch is given by the 
line integral of $l_a{\bf h}^a$ along the closed boundary curve.
Because ${\bf h}^a$ is exact, this integral vanishes. This is well known. 
To  see it explicitly when $D=2$,
express $\epsilon^{ab}$
in terms of the components of the unit tangent ${\bf t}=t^a {\bf e}_a$ and the normal
${\bf l}=l^a {\bf e}_a$, 
$\epsilon^{ab}= l^a t^b - l^b t^a$
so that
\begin{equation}
\oint ds \,l_a {\bf h}^a =
- \oint ds \,t^a \nabla_a {\bf A} \,.
\end{equation}
But $t^a\nabla_a = {\partial/\partial s}$ so that the closed line integral vanishes.
It would appear that the notion of a null stress is not very interesting;
as we will see this is not the case.

We are now in a position to comment on the behaviour of ${\bf f}_0^a$ under translations.
It was seen that 
under a translation ${\bf X}\to {\bf X} + {\bf a}$, ${\bf f}_0^a$ transforms by
${\bf f}_0^a\to
{\bf f}_0^a + {\bf h}_0^a$,where  ${\bf h}_0^a$ is
defined by Eq.(\ref{eq:f1def}).  It is easy to see that 
${\bf h}_0^a$ is, in fact, null.
Indeed, when $D=2$,
\begin{equation}
{\bf h}_0^a = 
\epsilon^{ab} {\bf a}\times {\bf e}_b
=\epsilon^{ab} \nabla_b [{\bf a}\times {\bf X}]\,,
\end{equation}
which is of the form
(\ref{eq:hdef}) with the identification,
${\bf A}= {\bf a}\times {\bf X}$.

While a change of origin has no physical significance, the value of $\bar{\bf f}^a$
at a point will depend on the choice made. It may be useful to
centre the origin to reflect the symmetry of  the configuration.
For example, in a sphere it would be perverse to place the origin anywhere 
but at the centre of the sphere unless there is a convincing reason to do otherwise. In the 
following section we will see that the explicit ${\bf X}$ dependence 
which appears in $\bar{\bf f}^a$ has a very important role.

\section{\sf An identity for rigid loops enclosing a fixed area}

A strong hint that casting the conservation law in terms of an effective
stress tensor might be useful is provided 
by examining a membrane which forms  an infinite cylinder. 
The cross-section of the cylinder perpendicular to its axis  
is a closed planar loop. The free energy for the cylinder 
ascribes rigidity to the loop. Its length $L$ is fixed and 
it encloses a fixed area $A$. Consider now a 
curve embedded on the plane parametrized by arclength $s$, ${\bf X}= {\bf X}(s)$; a prime
will denote a derivative with respect to $s$, so that ${\bf t}=
{\bf X}'$ is the unit tangent vector. The  curvature $k$ is
defined  by ${\bf t}' = - k {\bf n}$
and ${\bf n}' = k {\bf t}$,
where ${\bf n}$ is the unit normal. The energy penalty associated
with bending is given by 
\begin{equation}
H_0=\int ds\,  k^2\,.
\end{equation}
To facilitate comparison,
we will use the  same normalization for the bending energy 
introduced in \cite{ACCG} (without the factor of a half), 
but adopt a more conventional soft-matter
notation for the multipliers. 
Now the energy functional capturing the constraints on $L$ and $A$ is
\begin{equation}H= H_0 + \sigma L - PA\,.
\end{equation}
Consider the problem of determining the equilibria of this loop.
The Euler-Lagrange equation is
\begin{equation}
2 k'' + k^3 - \sigma k + P=0\,,
\end{equation}
and it has the first integral
\begin{equation}
{1\over 2}k'^2 + V(k) = E \,, 
\label{eq:Edef}
\end{equation}
where 
\begin{equation}
V(k) = {1\over 8} k^4 - {\sigma\over 4} k^2 + {P\over 2} k\,,
\end{equation}
and $E$ is a constant. The closure of the loop will quantize $E$. 
The quadrature determines the curvature $k$ implicitly as a function of $s$ 
in terms of elliptic functions. One would 
expect to require two integrations of the curvature to reconstruct the
corresponding loop; remarkably, none is needed. This was shown in \cite{ACCG}.
Here, it will be shown how this comes 
about using the conservation of the effective stress. 

The tension in the loop is given by
\begin{equation}
{\bf T} = (k^2 -\sigma) {\bf t} - 2 k'{\bf n}\,,
\end{equation}
and it satisfies
\begin{equation}
{\bf T}' =  P\, {\bf n}\,.
\label{eq:Tprn}
\end{equation}
For a planar curve ($D=1$), Eq.(\ref{eq:ndiv}) simplifies
${\bf n} = [({\bf X}\cdot {\bf t})\, {\bf n} - ({\bf X}\cdot {\bf
n})\, {\bf t} ]'$. The equilibrium condition Eq.(\ref{eq:Tprn}) may thus be recast as an 
identity for a space vector 
\begin{equation}
[{\bf X}\cdot {\bf n}+ P^{-1} (k^2 -\sigma)]\,  {\bf t}- 
[{\bf X}\cdot {\bf t} + 2 P^{-1} k']\,  {\bf n} = {\bf C}\,,
\label{eq:tn}
\end{equation}
where ${\bf C}$ is some constant vector along the loop. This constant is the analogue
of ${\bf h}^a$ discussed earlier in the two-dimensional context. However,
it is always possible to translate the origin so that ${\bf
X}\cdot {\bf t}=0$ when $k'=0$, which  gives ${\bf C}=0$. 
Eq.(\ref{eq:tn}) now provides two 
identities, one for each projection
\begin{eqnarray}
{\bf X}\cdot {\bf n} &=& P^{-1} (\sigma - k^2 )\nonumber\\
{\bf X}\cdot {\bf t} &=& - 2 P^{-1} k'\,.
\end{eqnarray}
The `$\sigma$' identities which were
identified in \cite{ACCG} are reproduced in a very direct way.
The latter of the two identities implies that
\begin{equation}
{\bf X}^2 =X_0^2 -  4P^{-1} k \,,
\label{eq:X2}
\end{equation}
where $X_0$ is a constant. Completeness 
of the basis vectors ${\bf t}$ and ${\bf n}$
together with Eq.(\ref{eq:Edef}) 
can now be used to express $X_0^2$ directly in terms of the 
constant $E$ appearing in the quadrature:
\begin{equation}
X_0^2 = P^{-2}(8E + \sigma^2)\,.
\end{equation}
Once the spectrum of $E$ has been determined, the 
curvature at any point detemines completely the loop geometry at that point. No integrations are 
necessary.  While it is well known that the Euler-Lagrange equation can be integrated 
to provide a quadrature for the curvature, this  
additional level of integrability is not evident in the Euler-Lagrange equation. 
In a beautiful paper, Joel Langer 
mentions the existence of such an identity somewhat obliquely as a property of the 
torsion-free limit of a certain integrable model for space curves \cite{Langer}.

It should be remarked that there is considerably greater freedom associated with an 
infinite or open elastic curve due to boundary conditions. 
No simple relationship between curvature and position is known. Nor is it expected.
The configuration will depend sensitively on the boundary conditions
(see, for example, \cite{Igor}). Isolation is important.

\section{\sf Vanishing effective stress implies spherical 
symmetry}

It is not unreasonable to inquire if there exist identities  between curvature and position
in solutions to $\nabla_a\bar{\bf f}^a=0$ analogous to those 
for the planar elastic loop. We begin with a search for 
solutions  satisfying the straightforward 
analogue to a vanishing ${\bf C}$ in Eq.(\ref{eq:tn}), $\bar {\bf f}^a =0$. 

Recall that translational invariance and reparametrization invariance together
imply a set of integrability conditions on the components of the stress \cite{Stress}. 
These conditions can be isolated explicitly by
expanding ${\bf f}^a$  with respect to the basis 
$\{{\bf e}_a,{\bf n}\}$
\begin{equation}
{\bf f}^{a} =   f^{ab}  {\bf e}{}_b + f^{a} {\bf n}\,.
\label{eq:ftn}
\end{equation}
Here $f^{ab}$ is a surface tensor; $f^a$ is a surface vector.
The projections of
Eq.(\ref{eq:Ef}) perpendicular and parallel to the surface 
then read
 \begin{eqnarray}
{\nabla}_a \,f^{a} -  K_{ab}{}  f^{ab}  &=& {\cal E}  \,,\label{eq:divperp}
\\
\nabla_a \, f^{ab} +  K^{b}{}_{a}   f^{a} &=& 0\,.\label{eq:divpar}
\end{eqnarray}
The
(normal) Euler-Lagrange derivative is captured completely
by the normal projection of the divergence; the projection process 
dismantles the natural surface divergence of ${\bf f}^a$.

The remaining $D$ identities (\ref{eq:divpar}) provide consistency 
conditions on the stress;
they are completely independent of ${\cal E}$; thus, they are also 
independent of the equilibrium. In particular,
if $H$ is a sum of terms, as it is for our constrained fluid membrane,
Eq.(\ref{eq:divpar}) will hold for each of them separately.
In Eq.(\ref{eq:divperp}), both the tangential stress $f^{ab}$ and the Euler-Lagrange  
derivative ${\cal E}$ act as a source for the normal stress. 
In Eq.(\ref{eq:divpar}), the normal stress provides a source for the
tangential stress. Neither stress
alone will generally be conserved. This coupling between normal and 
tangential stresses should be contrasted with the simple 
intrinsic conservation law for the stress $T^{ab}$ associated with a 
field coupling to the intrinsic surface geometry, $\nabla_a T^{ab}=0$.

Using Eq.(\ref{eq:fdef}),  
it is straightforward to identify the stress in a fluid membrane
with an energy given by Eq.(\ref{eq:Hfluid}). One finds for the 
surface tensor $f^{ab}$ and  
the vector $f^a$:
\begin{eqnarray}
f^{ab} &=&   K (K^{ab} - {K\over 2} g^{ab}) + \beta (K^{ab} - K
g^{ab})
-\sigma g^{ab}\nonumber\\
f^a &=& -   \nabla^a K \,. \nonumber
\end{eqnarray}
The corresponding Euler-Lagrange derivative is given by \cite{Stress,Hel.OuY:87}
\begin{equation}
{\cal E} =  -\, \nabla^2 K +
{1\over 2} K (K^2 - 2 K_{ab} K^{ab} ) + \beta {\cal R} +\sigma K
\,. \label{eq:elch}
\end{equation}
The intrinsic scalar curvature ${\cal R}$ is related to the extrinsic curvature 
by the contracted Gauss-Codazzi 
\begin{equation}
{\cal R}= K^2 - K^{ab} K_{ab}\,.
\label{eq:GC}
\end{equation}
This setup was described in
\cite{Stress} and refined in \cite{auxil}. Note the following points:

\vskip1pc
\noindent (1) The surface tensor $f^{ab}$ is
quadratic in $g_{ab}$ and $K_{ab}$.

\vskip1pc\noindent (2)
In a soap film, $f^{ab} = - \sigma g^{ab}$ is isotropic, with tension $\sigma$;
in general, however, the mechanical surface 
tension is not $\sigma$. The tangential stress $f^{ab}$ depends 
not only on $g_{ab}$ but also on the local state of bending
as characterized by $K_{ab}$.

\vskip1pc\noindent (3)
Because $f^{ab}$ depends locally on $K_{ab}$, its eigenvectors
coincide with those of $K_{ab}$; the corresponding eigenvalues are
quadratic functions of the principal curvatures.

\vskip1pc\noindent (4) Umbilical points are those where the principal curvatures 
coincide, so that   $K_{ab} =g_{ab} K/D$
(all points on a spherical membrane are umbilical).
At such a point (when $D=2$), the bending contribution to the stress vanishes.
What stresses are present are there because of competing constraints
on the membrane geometry.

\vskip1pc
The components of the effective stress are given by 
\begin{equation} 
\bar f^{ab} = 
f^{ab}  + {P\over D}
\, ({\bf n}\cdot {\bf X}) \, g^{ab}  \,,
\label{eq:ft0}
\end{equation}
and
\begin{equation}
\bar f^a = f^a
 - {P\over D} ({\bf e}^a\cdot {\bf X})\,.
\label{eq:fn0}
\end{equation}
Various interesting scalars can be constructed 
using the tangential effective stress $\bar f^{ab}$
given by (\ref{eq:ft0}). In the appendix, two of these 
scalars are shown to be surface divergences in equilibrium.

It is simple  to show that
$\bar {\bf f}^a=0$ in a spherical equilibrium when the origin is located at the centre
of the sphere. To confirm this, note that on a $D$-dimensional sphere of radius $R$,   $K=D/R$;
in addition, on a sphere every point is umbilical, $K_{ab}= K g_{ab} /D$, so that
the Euler-Lagrange equation ${\cal E}=P$ reduces to the vanishing 
of the following cubic polynomial in $R$: 
\begin{equation}
4\left(1 - {2\over D}\right)   + 2 \beta R +
\sigma R^2 - P R^3=0
\,. 
\label{eq:Rpoly}
\end{equation}
This equation degenerates into a quadratic when $D=2$ because of the scale invariance
of the Canham-Helfrich energy. 
Now, if the origin is placed at the centre of the sphere, ${\bf X}= R {\bf n}$.
It is straightforward to show that
$\bar f^{ab} =0 $ coincides with the identity Eq.(\ref{eq:Rpoly}). 
In addition, $\bar f^a =0$ because $K$ is constant
and ${\bf X}$ is orthogonal to the surface.
Thus $\bar {\bf f}^a= 0$. This is independent of the dimension. 
In particular, the scale invariance of the Canham-Helfrich energy when 
$D=2$ does not play a role. 

The only equilibrium configurations consistent with 
$\bar {\bf f}^a=0$ are, in fact, these spheres.
This follows from the vanishing of $\bar f^{ab}$. To see this, note that 
$\bar f^{ab}=0$ can always be cast in the form
\begin{equation}
K_{ab}= F\, g_{ab} \,,
\label{eq:KFg}
\end{equation}
where 
\begin{equation} F   = \left[ {K^2\over 2} + \beta  K + \sigma 
-  {P\over D}
\, ({\bf n}\cdot {\bf X})\right]\Bigg/ \Big[K+ \beta\Big]\,.
\label{eq:Fdef}
\end{equation}
In general, one can use completeness of the (orthonormal)
principal vectors $l_a^1$ and $l_2^a$ to provide a spectral expansion for 
$K^{ab}$ ($K_1$ and $K_2$ are the principal curvatures) 
\begin{equation}
K^{ab}= K_1 \, l^a_1 l^b_1 + K_2 \,
l^a_2 l^b_2 \,.
\label{eq:spec}
\end{equation}
Eq.(\ref{eq:KFg}) then implies that the principal curvatures must coincide everywhere:
$K_1=F=K_2$. What is more,  they must also be constant. The simplest way to see this 
is to appeal to the  contracted Codazzi-Mainardi equations, 
$\nabla_b K^{ab} -\nabla^a K=0$, which imply that $\nabla_a F=0$. Thus
$F$ must be constant on the surface. The only solution is therefore spherical. 
Note that there is no corresponding constraint on the curvature of a loop (when $D=1$ the
Codazzi-Mainardi equations are vacuous).

There are a few noteworthy alternative formulations of this 
non-existence result. Consider the following:
if $\bar f^a =0$ in equilibrium, the equilibrium must be
spherical and $\bar f^{ab}=0$.\footnote {\sf If $\bar f^a=0$, then 
$K_{ab} \bar f^{ab}=0$ must also hold when the Euler-Lagrange equation
is satisfied.}
Note that
if $\bar f^a=0$, where $\bar f^a$ is given by Eq.(\ref{eq:fn0}),
then 
\begin{equation}
K= C- {1\over 2D} P {\bf X}^2
\,,
\label{eq:KCP}
\end{equation}
where $C$ is a constant. This looks like a very direct analogue of Eq.(\ref{eq:X2}). 
It turns out, however, that this equation is generally inconsisitent with 
its tangential counterpart; indeed the only equilibrium consistent with it is spherical.
For if $\bar f^a=0$ then
$\nabla_a \bar f^{ab}=0$ in equilibrium. But, on substituting the 
the expresssion for $\bar f^{ab}$ given by Eq.(\ref{eq:ft0}),  this statement reads
\begin{equation}
K^{ab} [\nabla_a K  - {P\over 2} ( {\bf e}_a  \cdot {\bf X}) ] =0\,.
\end{equation}
If Eq.(\ref{eq:KCP}) is now used to express $K$ in terms of ${\bf X}$, there follows
${\bf X}\cdot {\bf e}_a=0$ or ${\bf X}^2 $ is constant.
Spherically symmetry is forced on us.
In the next section we will explain why
the naive analogue of the loop identities comes up short and suggest a 
generalization.

\section{\sf Is there a generalization of the loop identities?}
\label{generalize}

It has been shown that a direct analogue of the
loop identities does not exist unless the geometry is spherical.
There is, however, a good reason for this  failure:
on the curve, the only null stress is 
constant; on a two-dimensional surface, however,
any space vector field will generate a null stress.
A more appropriate analogue is an identity of the form,
$\bar {\bf f}^a = {\bf h}^a$, 
where ${\bf h}^a$ is some (hopefully) local null stress.
Let us express ${\bf h}^a= h^{ab} {\bf e}_a + h^a {\bf n}$.
In components,
\begin{equation} f^{ab} + {P\over D}
\, ({\bf n}\cdot {\bf X}) \, g^{ab} = h^{ab}\,,
\label{eq:fh1}\end{equation}
and
\begin{equation}
f^a - {P\over D} ({\bf e}^a\cdot {\bf X})= h^a\,.
\label{eq:fh2}
\end{equation}
To consistently couple $h^{ab}$ to a symmetric $f^{ab}$,  
$h^{ab}$ must also be symmetric. On one hand, 
 this symmetry  does not itself possess an invariant significance. This is because a 
null addition of the form (\ref{eq:hdef})  will not generally
have a symmetric tangential component. However, the 
canonical $\bar {\bf f}^a$ for a fluid membrane does come
with this symmetry so  if we are to couple ${\bf h}^a$ consistently to
${\bf f}^a$, it had better also possess it.

It is simple to characterize null potentials ${\bf A}$ (defined in 
section \ref{sec:amb}) which yield
symmetric $h^{ab}$. In appendix (\ref{app:areapres}) we show that for $D=2$,  
$h^{ab}$ is symmetric when $\nabla_a {\bf A}\cdot {\bf e}^a=0$.
This identifies the vector potential ${\bf A}$ 
with an area preserving deformation of the surface.
It is straightforward to construct null stresses
of this form. They are also of considerable interest in their 
own right leading to unexpected connections between geometrical 
invariants. Work in this direction will be reported elsewhere.

Whereas the symmetry of $h^{ab}$ is a necessary condition on ${\bf h}^a$,
it is clearly not a sufficient one. Indeed, it is straigtforward to 
construct a counterexample. Suppose that $h^{ab}$ is a linear
combination of $g^{ab}$ and $K^{ab}$. The null stress ${\bf h}^a$ 
derived from the potential ${\bf A} =
{\bf X} \times {\bf n}$ is of this form. It is then clear from
Eq.(\ref{eq:fh1}) that $K^{ab}$ must be proportional to $g^{ab}$. 
An identical argument to the one employed earlier kills all but the spherical solution. 
Unfortunately, we are not even close to an exhaustive list 
of necessary and sufficient conditions on ${\bf h}^a$.  
It will clearly be necessary to introduce some element of anisotropy into
${\bf h}^a$ if spherical symmetry is to be broken: its 
symmetry will reflect the symmetry of 
the configuration.

It is not, of course, obvious that Eqs.(\ref{eq:fh1}) and (\ref{eq:fh2}) are going  
to be tractable; they are very different from the Euler-Lagrange equations
we are familiar with. What is clear is that it is worth finding out;
for, if they do turn out to be tractable, the implications will be
far-reaching.

\section{\sf Discussion}

We have shown that the Laplace pressure on a closed vesicle enforcing the constraint 
on the enclosed volume can be incorporated into a 
conserved effective stress tensor. What might appear to be some
mathematical sleight-of-hand does turn out to lend insight 
into the nature of the surface geometry. 

What is perhaps most intriguing about this set-up is 
that it suggests that, in an isolated vesicle, it may be possible to integrate
the conservation law. In a loop, where this is straightforward to do,
it leads to powerful identities between the curvature 
and the position; control is provided over curvature without any need to solve
the Euler-Lagrange equation itself explicitly.  The generalization to higher-dimensional 
surfaces, involving the identification of the effective stress tensor with an appropriate null stress,
is unfortunately anything but straightforward. 
The challenge is to implement it.

It should be pointed out that our results can also be cast  
in the language of differential forms (see, for example, 
\cite{IveyBryant}, and in the specific context of membranes \cite{Tu}). This is also likely  
to be useful.

\vskip3pc
\noindent{\large \sf Acknowledgments}

\vspace{.5cm}

Thanks to Riccardo Capovilla, Pavel Castro-Villareal and  Chryssomalis Chryssomalakos in Mexico City
as well as  Markus Deserno and Martin M\"uller in Mainz for comments.
Partial support from CONACyT grant 44974-F is acknowledged.

\appendix

\section{\sf Derivation of Eq.(\ref{eq:delV1})}

To derive Eq.(\ref{eq:delV1}), recall that the
deformation $\delta {\bf X}$ induces the change $\delta g_{ab}= {\bf e}_{a}\cdot
\nabla_{b}\delta {\bf X} + 
{\bf e}_{b}\cdot
\nabla_{a}\delta {\bf X}$ in the metric, so that $\delta dA = dA \,{\bf
e}^a\cdot\nabla_a\delta {\bf X}$.  
The deformation in the unit
normal is tangential, $\delta {\bf n}= -({\bf n}\cdot\nabla_a
\delta{\bf X})\, {\bf e}^a$. Thus
\begin{eqnarray}
\delta V &=&
{1\over D+1}\int dA \, \left[({\bf n}\cdot {\bf X})\,({\bf e}^a\cdot
\nabla_a\delta {\bf X})
- ({\bf e}^a\cdot {\bf X})\,({\bf n}\cdot \nabla_a \delta{\bf X})
+ ({\bf n}\cdot \delta {\bf X})\right]\nonumber\\
&=&
{1\over D+1} \int dA \, \left[({\bf n}\cdot \delta {\bf X})
-  ({\bf f}_0^a \cdot  \nabla_a \delta{\bf X})\right]\nonumber\\
&=&
{1\over D+1} \int dA \, \left[({\bf n}\cdot \delta {\bf X})
+  (\nabla_a {\bf f}_0^a \cdot\delta{\bf X})\right]
- {1\over D+1}\int   dA\,  \nabla_a ({\bf f}_0^a \cdot\delta{\bf X})
\,,
\label{eq:delV2}
\end{eqnarray}
where Eq.(\ref{eq:f0def}) has been used on the second line
and all derivatives of $\delta {\bf X}$ have been collected in a
divergence on the third. 
Now use the identity Eq.(\ref{eq:ndiv})
on the second bulk term. Eq.(\ref{eq:delV1}) follows.

\section{\sf Identities involving $\epsilon^{ab}$}

For two-dimensional surfaces,
it is useful to express ${\bf f}_0^a$ as a double cross product:
\begin{equation}
{\bf f}_0^a = {\bf X}\times ({\bf n}\times {\bf e}^a) = 
\epsilon^{ab} \, {\bf X}\times {\bf e}_b\,.
\end{equation}
Here 
$\epsilon^{ab}$ is the two-dimensional
Levi-Civita antisymmetric tensor.
We have ${\bf e}_a \times {\bf e}_b = \epsilon_{ab} {\bf n}$, so that
${\bf n}\times {\bf e}^a = \epsilon^{ab} {\bf e}_b$.
We note that 
$\epsilon^{ab}\epsilon^{cd}= g^{ac} g^{bd} - g^{ad} g^{bc}$;
and, as a consequence, $\epsilon^{ac} \epsilon_{c}{}^b = - g^{ab}$.

\section{\sf Two scalars}

We examine various scalars that can be constructed 
using the tangential effective stress $\bar f^{ab}$
given by (\ref{eq:ft0}). 

The trace $g_{ab} \bar f^{ab}$ is given by
\begin{equation}
  g_{ab}\bar f^{ab} = \left(1-{D\over 2}\right)\, K^2 +
(1-D) \beta  K - D\sigma  +  P \,({\bf n}\cdot {\bf X})\,.
\label{eq:bar1}
\end{equation}
In general,  $g_{ab} \bar f^{ab}$ will be  a divergence in equilibrium.
This is a consequence of the fact, observed in \cite{Stress}, that
 \begin{equation}
  g_{ab}\bar f^{ab} = \nabla_a ( {\bf X}\cdot \bar {\bf f}^a)\,,
\label{eq:trdiv}
\end{equation}
when $\nabla_a \bar {\bf f}^a=0$.
This identifies the argument of the divergence as 
${\bf X}\cdot \bar {\bf f}^a$. Upon integration of 
Eq.(\ref{eq:bar1}) over the closed surface, there follows 
\begin{equation}
 (2-D) H_0 +
\left(1-D\right) \beta  M - D\sigma A +  (D+1) P V =0\,.
\label{eq:virial}
\end{equation}
When $D=2$, this is the {\it well-known} virial identity
associated with the scaling behaviour of the energy in equilibrium
\cite{Reviews}.
We also note that the vanishing of $g_{ab}\bar f^{ab}$ is consistent with 
this identity.

In addition, $K_{ab} \bar f^{ab}$ is a divergence. This follows from 
the equilibrium condition,
$\nabla_a \bar f^a + K_{ab} \bar f^{ab} =0$. 
We have 
\begin{equation}
K_{ab} \bar f^{ab} =
{1\over2} K (K^2 - 2 {\cal R}) - \beta {\cal R} - \sigma K +
{P \over D} K\,({\bf n}\cdot {\bf X})\,.
\label{eq:bar2}
\end{equation}
The contracted 
Gauss-Codazzi equation (\ref{eq:GC})
has been used to express the  quadratic $K_{ab}K^{ab}$ in 
terms of ${\cal R}$ and $K^2$. 
Using the Minkowski identity for the area \cite{Spivak},
\begin{equation}
A= {1\over D}\int dA\, K\,({\bf n}\cdot {\bf X})\,,
\end{equation}
one sees that the identity
\begin{equation}
\int dA \,  K_{ab} \bar f^{ab} =0 
\end{equation}
is consistent with the integrated Euler-Lagrange equation with
${\cal E}$ given by Eq.(\ref{eq:elch}).

\section{\sf Symmetric $h^{ab}$ and 
area-preserving deformations}
\label{app:areapres}

Note that $h^{ab}= {\bf h}^a\cdot {\bf e}^b$ and
$h^{ab}$ will be symmetric if $\epsilon_{ab} h^{ab}= 0$. We now use 
Eq.(\ref{eq:hdef}) to express this condition  in terms of a potential:  
\begin{equation}
\epsilon_{ab} h^{ab}= 
\nabla_a {\bf A}\cdot {\bf e}^a\,.
\end{equation}
Thus $h^{ab}$ is symmetric when $\nabla_a {\bf A}\cdot {\bf e}^a=0$. 
But this condition identifies the vector potential ${\bf A}$ 
with an area preserving deformation of the surface: Under the deformation
$\delta {\bf X}= {\bf A}$ , it follows from the definition of $g_{ab}$ that
$\delta g_{ab}= {\bf e}_{a}\cdot
\nabla_{b}{\bf A} + {\bf e}_{b}\cdot
\nabla_{a} {\bf A}$. The area element is proportional to $\sqrt{g}$, where 
$g= {\rm det}\,g_{ab}$. However,
\begin{equation}
\delta g = g g^{ab} \delta g_{ab} =
2 g {\bf e}^a\cdot \nabla_a {\bf A}\,.
\end{equation}
Thus $\delta \sqrt{g}=0$ when $\nabla_a {\bf A}\cdot {\bf e}^a=0$. 
This condition can  be cast 
in terms of components tangent and normal to the surface, 
${\bf A}= A_{\|}^a {\bf e}_a + A_\perp {\bf n}$. Then 
 $\nabla_a {\bf A}\,\cdot\, {\bf e}^a=0$ is equivalent to 
$\nabla_a A_{||}^a + K A_\perp =0$.


\begin{thebibliography}{99}

\bibitem{Kenmotsu} A nice discussion of this 
problem is provided in 
Kenmotsu K {\it Surfaces with constant mean curvature} 
(American Mathematical Society, Providence Rhode Island 2003)

\bibitem{CanHel} Canham P 1970   {\it J. Theor. Biol.} {\bf 26} 61;
Helfrich W 1973 {\it Z. Naturforsch.} {\bf C28} 693


\bibitem{Willmore} Willmore TJ, {\sl Total Curvature in Riemannian
Geometry} (Chichester: Ellis Horwood, 1982)

\bibitem{SackLip} Lipowsky R and Sackmann E
{\it {S}tructure and {D}ynamics of {M}embranes, {V}ol. 1 and 2}
(Handbook of Biological Physics, Elsevier Science B.V. 1995)

\bibitem{Reviews} 
Seifert U 1997 {\it Adv. in Phys.} {\bf 46} 13

\bibitem{Nonmed} Svetina S and 
\u{Z}ek\u{s} B 1996 in
{\it Nonmedical Applications of Liposomes}, eds. D.D. Lasic and
Y. Barenholz (CRC: Boca Raton, FL)

\bibitem{Wor} 
Lim G, Wortis M and Mukhopadhyay R 2002
{\it PNAS} {\bf 99} 16766

\bibitem{Zit}
Ziherl P and Svetina S
{\it Europhys. Lett.} 2005 {\bf 70}  690

\bibitem{Evans1} Evans E and Skalak R {\it  Mechanics and Thermodynamics
of Biomembranes} (CRC Press., Boca Raton, 1980)

\bibitem{Kral} 
Kralchevsky PA Eriksson JC and Ljunggren S 1994 {\it Adv.
Coll. and Interface. Sci.} {\bf 48} 19 

\bibitem{Stress} Capovilla R and 
Guven J 2002 {\it J. Phys. A: Math. and Gen.} {\bf 35} 6233

\bibitem{auxil} Guven J 2004
{\it J. Phys. A: Math and Gen.} {\bf 37} L313

\bibitem{LomMiao}  Lomholt MA and Miao L
{\it Two different descriptive approaches to the mechanics of membranes and the connection 
between them} cond-mat/0509664

\bibitem{Ruso} 
Novozhilov VV {\it Thin shell theory, Second Edition} (Groningen: Noordhoff 1964)


\bibitem{Sve.Zek:89} Svetina S and \u{Z}ek\u{s} B 1989 {\it Eur. Biophys.
J.}  {\bf 17} 101

\bibitem{ACCG}  
Arreaga G, Capovilla R, Chryssomalakos C and Guven J 2002
{\it Phys. Rev. E} {\bf 65} 031801;
Capovilla R, Chryssomalakos C and Guven J 2002
{\it E. Phys. J.} {\bf B29} 163

\bibitem{Wulff} Pimpinelli A and Villain J {\it Physics of crystal growth}
(Cambridge University Press 1998) 

\bibitem{Carmo} Do Carmo M {\it Differential geometry of curves and surfaces}
(Prentice Hall 1976)

\bibitem{Spivak} Spivak M {\it A Comprehensive Introduction to
Differential Geometry. Vols. 1-5, Second Edition} (Publish or
Perish, 1979)

\bibitem{Deserno} Deserno M 2005 {\it Lecture Notes on Differential Geometry}

\bibitem{JG05} Guven J 2005 {\it J. Phys. A: Math and Gen.} {\bf 38}  7943


\bibitem{MDG} M\"uller M, Deserno M and Guven J 
{\it Euro. Phys. Lett.} 69 (2005) 482-488;
{\it Phys. Rev. E} {\bf 72} (2005)  061407  
`Interface mediated interactions between particles ---
a geometrical approach' arXiv:cond-mat/0506019

\bibitem{Flanders} Flanders H 
{\it Differential forms with applications to the physical sciences}
(Dover Publications Inc. New York 1989)

\bibitem{Langer} Langer J 1999 {\it New York J. Math} {\bf 5} 25 

\bibitem{Igor}
Kulic IM 2004 
{\it Statistical Mechanics of Protein Complexed and Condensed DNA}
(Ph.D. thesis, Max-Planck-Institut for Polymer Research, Mainz)

\bibitem{Hel.OuY:87}
Zhong-Can OY and Helfrich W 1987 {\it Phys. Rev. Lett.} {\bf 59}
2486; 1989 {\it Phys. Rev. A} {\bf 39} 5280 


\bibitem{IveyBryant} Ivey TA and Landsberg JM {\it Cartan for beginners} 
(American Mathematical Society, Providence Rhode Island 2003);
Griffiths P et al 2003
Exterior Differential Systems and Euler-Lagrange Partial
Differential Equations (Chicago Lectures in Mathematics Series)

\bibitem{Tu}
Tu ZC and Ou-Yang ZC 2004 {\it J. Phys. A: Math. Gen.} {\bf 37} 11407

\end{thebibliography}
\end{document}